\documentclass[twocolumn]{aastex631}
\usepackage{amsmath}

\shorttitle{sBHs can stretch accretion disks}
\shortauthors{Zhou et al.}

\begin{document}

\title{Stellar black holes can ``stretch'' supermassive black hole accretion disks}

\correspondingauthor{Mouyuan Sun}
\email{msun88@xmu.edu.cn}

\author[0009-0005-2801-6594]{Shuying Zhou}
\affiliation{Department of Astronomy, Xiamen University, Xiamen, Fujian 361005, China; msun88@xmu.edu.cn}

\author[0000-0002-0771-2153]{Mouyuan Sun}
\affiliation{Department of Astronomy, Xiamen University, Xiamen, Fujian 361005, China; msun88@xmu.edu.cn}

\author[0000-0001-8678-6291]{Tong Liu}
\affiliation{Department of Astronomy, Xiamen University, Xiamen, Fujian 361005, China; msun88@xmu.edu.cn}

\author[0000-0001-9449-9268]{Jian-Min Wang}
\affiliation{Key Laboratory for Particle Astrophysics, Institute of High Energy Physics, Chinese Academy of Sciences, 19B Yuquan Road, Beijing 100049, China}
\affiliation{School of Astronomy and Space Sciences, University of Chinese Academy of Sciences, 19A Yuquan Road, Beijing 100049, China}
\affiliation{National Astronomical Observatory of China, 20A Datun Road, Beijing 100020, China}

\author[0000-0002-4419-6434]{Jun-Xian Wang}
\affiliation{Department of Astronomy, University of Science and Technology of China, Hefei 230026, China}
\affiliation{School of Astronomy and Space Science, University of Science and Technology of China, Hefei 230026, China}

\author[0000-0002-1935-8104]{Yongquan Xue}
\affiliation{Department of Astronomy, University of Science and Technology of China, Hefei 230026, China}
\affiliation{School of Astronomy and Space Science, University of Science and Technology of China, Hefei 230026, China}

\begin{abstract}
Stellar black holes (sBHs) are widely believed to exist in the accretion disks of active galactic nuclei (AGNs). Previous studies often focus on the transient emission produced by embedded sBHs. Here, we explore the possible observational consequences of an AGN accretion disk that contains a population of accreting sBHs. Embedded accreting sBHs change the effective temperature distribution of the AGN accretion disk by heating gas in the outer regions. Two possible observational consequences are presented. First, the spectral energy distribution has a turnover feature at $\sim 4700\ \textrm{\AA}$ when the supermassive black hole (SMBH) mass is $\sim 10^8\ M_{\odot}$, which can help explain the observed shallow spectral shape at wavelengths $>5000\ \textrm{\AA}$ for the Sloan Digital Sky Survey quasar composite spectrum. Second, the half-light radius of a given relatively long wavelength is significantly larger than for an AGN disk without sBHs, which can be tested by microlensing observations. With appropriate sBH distributions, the model can be reconciled with quasar microlensing disk sizes. We propose that the half-light radius-wavelength relation can be utilized to investigate the distributions of embedded sBHs in AGN accretion disks.
\end{abstract}


\section{Introduction} \label{sec:intro}

Star formation in the outer regions of the accretion disks of Active Galactic Nuclei (AGNs) is a widely studied topic \citep[e.g.,][]{Shlosman1989, Collin1999, Goodman2004, Wang2011, Cantiello2021, Wang2023}. The outer regions of the accretion disk are far from the central supermassive black hole (SMBH), where self-gravity plays a dominant role, and the disk inevitably collapses to form stars \citep[e.g.,][]{Kolykhalov1980, Shlosman1989}. Nuclear star formation can also be triggered by jetted tidal disruption events in SMBHs \citep[e.g.,][]{Perna2022}. In addition, stars near the center of the galaxy can interact with the SMBH accretion disk, resulting in a loss of orbital energy and angular momentum, and are captured by the accretion disk \citep[e.g.,][]{Syer1991, Cantiello2021}. The redshift-independent high metal abundance in quasar broad-line regions provides strong observational evidence for star formation activities in SMBH accretion disks \citep[e.g.,][]{Artymowicz1993, Hamann1999, Wang2011, Qi2022, WangS2022}. Stars in accretion disks can accrete gas and rapidly evolve into compact objects, such as white dwarfs, neutron stars, and stellar black holes (sBHs), as first pointed out by \cite{Cheng1999}. 

The discovery of the Fermi bubbles strongly suggests that Sagittarius A$^{*}$ was an active AGN in the past \citep{Su2010,Yang2022NatAs}. During this active stage, stars or compact objects may form in the accretion disk of Sagittarius A$^{*}$. Although the SMBH accretion in Sagittarius A$^{*}$ is now quenched, the present-day stellar and compact object populations should be affected by star formation epochs during the past active phase of Sagittarius A$^{*}$. Very recently, \cite{Gravity_Collaboration2023} find that the Sagittarius A$^{*}$ flare is located at about nine gravitational radii, consistent with the Keplerian orbital motion of the innermost accretion zone hot spot. While the physical origin of the flare remains unclear, it is proposed that a star \citep{Leibowitz2020} or a $\sim 40\ M_{\odot}$ sBH \citep{Wang2023_SgrA} orbiting periodically around the central SMBH drives this Sagittarius A$^{*}$ flare through interaction with SMBH accretion disk. The star or the sBH is embedded in the advection-dominated accretion flow around the SMBH in the Galactic Center. 

In the special environment of the AGN accretion disk, stars and compact objects are prone to events such as massive stellar explosions and mergers or collisions of compact objects. These events can produce transient emission such as core-collapse supernovae \citep[e.g.,][]{Grishin2021}, $\gamma$-ray bursts \citep[e.g.,][]{Cheng1999, Ray2023}, neutrino bursts \citep[e.g.,][]{Zhu2021}, tidal disruption events \citep[e.g.,][]{Yang2022}, Bondi explosions \citep[e.g.,][]{Wang2021-Bondi} and binary black hole mergers \citep[e.g.,][]{Graham2020}. Mergers of binary black holes, black hole-neutron star systems, or binary neutron stars are important sources of gravitational waves detected by LIGO/Virgo/KAGRA \citep{Abbott2019}. GW190521 \citep{Abbott2020} is of considerable interest because the Zwicky Transient Facility detected its potential electromagnetic counterpart in AGN J124942.3+344929 \citep{Graham2020}. This implies that the binary black hole merger may occur in the AGN accretion disk \citep[but see][for the discussion of an association by chance]{Palmese2021}, which has been used to infer cosmological parameters \citep[e.g.,][]{Chen2022}. 

Although there have been many studies of sBHs in AGN accretion disks, we remain poorly informed about their number densities and distributions. Stars \citep[e.g.,][]{Thompson2005} or sBHs \citep[e.g.,][]{Gilbaum2022} in the self-gravity regions of AGN accretion disks can substantially increase the central temperature of gas. Here we show that sBHs embedded in the static standard disk \citep[hereafter SSD;][]{SSD} modify the effective temperature distribution of the accretion disk and change the disk half-light radii at various wavelengths. Different sBH distributions in the accretion disk have different degrees of influence on the half-light radius, which may be used to infer sBH distributions. 

Quasar microlensing observations provide a unique opportunity to resolve AGN accretion disks. Distant quasars might be gravitationally lensed by foreground galaxies and have multiple lensed images. There are time delays in the flux variations of different lensed images which can be used to constrain the cosmological model \citep[i.e., the strong lensing time-delay cosmography; for a review, see, e.g.,][]{Treu2016}. On top of the strong lensing effects, stellar objects (in the foreground galaxy) that might happen to be in the line of sight of a lensed quasar image can cause additional flux magnifications. These microlensing induced apparent flux variations are size dependent and can be used to effectively measure the half-light radii of AGN accretion disks \citep{Mortonson2005}. A number of studies have measured the half-light radii of AGN accretion disks using the microlensing effects; the results suggest that the measured half-light radii are 2--4 times larger than the SSD predictions \citep[e.g.,][]{Morgan2010}. 

In this work, we propose that quasar microlensing observations can be used to probe sBHs in AGN accretion disks. The manuscript is organized as follows. In Section~\ref{sec:Model}, we detail the model calculations; in Section~\ref{sec:results}, we present the spectral energy distribution of the model and the half-light radius as a function of wavelength; in Section~\ref{sec:discussion}, the implications for the probe of the sBH distribution, the AGN accretion physics, and strong lensing time-delay cosmography are discussed. The main conclusions are summarized in Section~\ref{sec:Summary}.

\section{Model} \label{sec:Model}
A large number of sBHs can form in situ or be captured from nuclear star clusters. For instance, \citet{Artymowicz1993} and \citet{Rozyczka1995} point out that the AGN accretion disk captures stars from nuclear star clusters at a rate of $10^{-4}$--$10^{-3}\ \mathrm{year}^{-1}$. The typical AGN lifetime is $10^{7}$--$10^{8}\ \mathrm{years}$. Thus, the total captured number of stars, which can further accrete gas and evolve into sBHs, is $10^{3}$--$10^{5}$ within the AGN lifetime. A detailed calculation of \cite{Gilbaum2022} by considering in situ formation, capture, and migration yields that the number of sBHs within a disk radius of $3-3000\ R_\mathrm{s}$ over the AGN lifetime can be $10^{2}$--$10^{8}$ (Figure 12 in their Section 3.2). Thus, we establish a model with $10^{3}$--$10^{5}$ \citep[e.g.,][]{Artymowicz1993, Rozyczka1995} sBHs embedded in the AGN accretion disk. Neutron stars may outnumber sBHs in the AGN accretion disk, but their energy outputs are smaller than sBHs; white dwarfs are difficult to form within the AGN lifetime because the typical evolutionary time of a white dwarf is $\sim 1$ billion years \citep[e.g.,][]{Catal2008}. Therefore, we only consider the effects of sBHs. We also assume that sBHs do not significantly affect the angular momentum transport in the AGN accretion disk \citep{Gilbaum2022}. The SMBH mass is significantly larger than the total mass of the sBHs, and the sBHs only play an essential role for the gas in their vicinity. A small fraction of the gas ($f_{\mathrm{sBHs}}$) in the AGN accretion disk is accreted by the sBHs rather than the central SMBH. This causes the effective temperature distribution of the AGN accretion disk to be different from the canonical $T_\mathrm{eff}\propto R^{-3/4}$ of the SSD theory, and the contributions of the luminosities at different radii to the total luminosity to be altered for a given wavelength. 

When embedded in the AGN accretion disk, sBHs can rapidly accrete gas from the AGN accretion disk. The total accretion rate for the embedded sBHs is $f_{\mathrm{sBHs}}\dot{M}_{\mathrm{tot}}$, where $\dot{M}_{\mathrm{tot}}$ is the total mass accretion rate in the AGN accretion disk. Each accreting sBH can produce intensive emission, preferably in the X-ray band. Given that the AGN disk is optically thick, these X-ray photons cannot escape freely but are absorbed by the AGN disk and heat the ambient gas. In this case, the AGN disk has two sources of heating: the local viscous heating in the AGN disk and the sBHs. We discuss in Section \ref{subsec:4.2} the possible consequence if sBHs are distributed in optically thin accretion-disk regions \citep[$\sim 10^4\ R_\mathrm{S}$;][]{Thompson2005}. We do not consider the possible interaction between the SMBH and sBH accretion processes. We argue this may be a good approximation if $f_{\mathrm{sBHs}}\ll 1$. 

The heating rates can be estimated as follows. For the local viscous heating in the AGN disk, the heating rate per unit area is 
\begin{equation} \label{equ:qvis}
    Q^{+}_{\mathrm{vis}}=\frac{3G M_{\bullet}\dot{M}_{\bullet}}{8\pi R^3}f_r \\,
\end{equation}
where $G$, $M_{\bullet}$, and $R$ are the gravitational constant, the SMBH mass, and the radius of the AGN accretion disk, respectively; the factor $f_r=1-(3R_\mathrm{S}/R)^{1/2}$, where $R_\mathrm{S}=2GM_{\bullet}/c^2$ is the Schwarzschild radius of the SMBH; the mass accretion rate to the SMBH $\dot{M}_{\bullet}=(1-f_{\mathrm{sBHs}})\dot{M}_{\mathrm{tot}}$. Meanwhile, the total heating rate due to sBHs per unit area depends upon their distribution on the AGN accretion disk. We divide the accretion disk from the inner boundary $R_\mathrm{in}$ to the outer boundary $R_\mathrm{out}$ into 256 equal rings on a logarithmic scale. For simplification, it is straightforward to assume that sBHs are equally densely distributed in each ring; the sBH number on each ring $N(R)$ is 
\begin{equation} \label{equ:NR}
    N(R)=2\pi R\Delta R\frac{N_\mathrm{sBHs}}{\pi(R^2_\mathrm{out}-R^2_\mathrm{in})},
\end{equation}
where $N_\mathrm{sBHs}$ is the total number of sBHs in the AGN accretion disk. As a result, most accreting sBHs reside in the outer regions of the AGN accretion disk. Note that real sBH distributions can be more complicated than this, as we discuss in Section \ref{subsec:4.1}. Each ring is further divided into 256 equal zones along the azimuthal direction, i.e., $\Delta \phi = 2\pi/256$. Hence, the size of each zone is comparable to the scale height of the AGN accretion disk. The sBHs in each zone will then be able to heat the gas within the zone isotropically. The heating rate in each zone is simply 
\begin{equation}\label{equ:qsbh}
    Q^{+}_{\mathrm{sBHs}} = N(R,\phi)L_{\mathrm{sBH}} \\,
\end{equation}
where $N(R,\phi)$ and $L_{\mathrm{sBH}}$ are the number of sBHs in a zone and the bolometric luminosity of each accreting sBH. We assume that all sBHs accrete at the Eddington limit\footnote{If the accretion rates of sBHs are not constrained to the Eddington limit, sBHs will grow rapidly into intermediate-mass black holes and accrete most gas that should flow into the SMBH, which may cause self-consistency problems in AGN accretion disk models \citep{Gilbaum2022}.}, and $L_{\mathrm{sBH}}$ equals to the Eddington luminosity of an sBH, which is $L_\mathrm{sBH,Edd}=1.26\times10^{38}M_\mathrm{sBH}/M_\mathrm{\odot}\ \mathrm{[erg\ s^{-1}]}$, where $M_\mathrm{sBH}$ and $M_\mathrm{\odot}$ are the sBH mass and the solar mass, respectively. Here, the radiative efficiency of $10\%$ is assumed. We stress that the radiative efficiency and the additional heating due to sBHs increase with the black-hole spin. Note that the Bondi accretion rate of a typical sBH in the AGN accretion disk is much larger than the Eddington accretion rate \citep[e.g.,][]{Wang2021-Bondi}. The real accretion process is more complex than the ideal Bondi case, and accompanying outflows may significantly modulate the accretion rate \citep[e.g.,][]{Takeo2020}. Hence, following \cite{Gilbaum2022}, we assume that sBHs accrete at the Eddington limit. Then, the total number of sBHs is
\begin{equation} \label{equ:sBHs_number}
    N_\mathrm{sBHs}=\frac{f_\mathrm{sBHs}\dot{M}_\mathrm{tot}}{\dot{M}_\mathrm{sBH,Edd}},
\end{equation}
where $\dot{M}_\mathrm{sBH,Edd}$ is the Eddington accretion rate of an sBH, i.e., $\dot{M}_\mathrm{sBH,Edd}=10L_\mathrm{sBH,Edd}/c^2$. 

\begin{figure*}
    \centering
    \includegraphics[width=8cm,height=8cm,keepaspectratio]{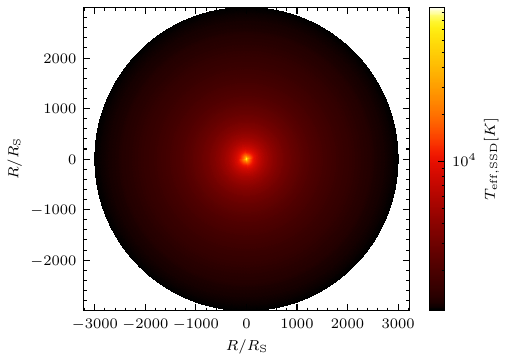}
    \includegraphics[width=8cm,height=8cm,keepaspectratio]{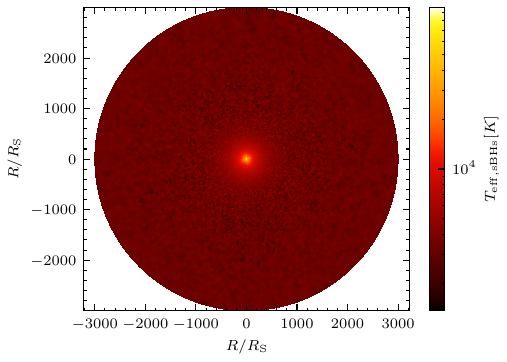}
    \includegraphics[width=7cm,height=7cm,keepaspectratio]{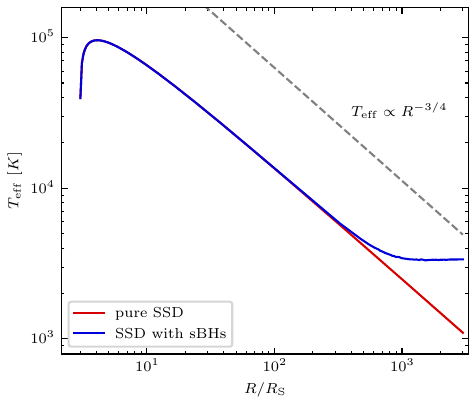}
    \caption{The effective temperature ($T_\mathrm{eff}$) distributions from $R_\mathrm{in}=3R_\mathrm{S}$ to $R_\mathrm{out}=3000R_\mathrm{S}$ for $M_{\bullet}=10^8\ M_{\odot}$ and $\dot{m}_{\mathrm{tot}}=\dot{M}_{\mathrm{tot}}/\dot{M}_\mathrm{\bullet,Edd}=0.3$. The top left panel is for a pure SSD; the top right panel is for an SSD with sBHs for $f_\mathrm{sBHs}=0.1$ and $M_\mathrm{sBH}=50\ M_{\odot}$ (Eq.~\ref{equ:Teff-sBHs}). The bottom panel compares the effective temperature profiles for the two cases. The red curve is the effective temperature profile for a pure SSD; the blue curve is the azimuthal averaged effective temperature profile for an SSD embedded with sBHs; the gray dashed line indicates the $T_\mathrm{eff}\propto R^{-3/4}$ scaling relation. The accreting sBHs in an SSD heat up gas in the outer regions, resulting in an effective temperature profile significantly shallower than a pure SSD. Microlensing observations can distinguish the two temperature profiles.}
    \label{fig1}
\end{figure*}

The heating due to sBHs plays an important role in the outer disk zones. Indeed, it is evident that the ratio 
\begin{equation} \label{equ:ratio}
    \epsilon = Q^{+}_{\mathrm{sBHs}}/(Q^{+}_{\mathrm{vis}} \Delta S(R,\phi))\propto (R/R_\mathrm{S})^3,
\end{equation}
where $\Delta S(R,\phi)=R\Delta\phi\Delta R$ is the area of each zone. For the case of sBHs embedded in an AGN accretion disk, the effective temperature profile of the AGN disk can be obtained by considering the balance between the heating rate and the disk surface cooling rate. We consider perfect blackbody radiation at each radius. Hence, the effective temperature $T_{\mathrm{eff,sBHs}}$ is  
\begin{equation} \label{equ:Teff-sBHs}
    2\sigma T_{\mathrm{eff,sBHs}}^4(R,\phi) = Q^{+}_{\mathrm{vis}} ( 1 + \epsilon )\\,
\end{equation}
where $\sigma$ is the Stefan-Boltzmann constant. We set $R_\mathrm{in} = 3\ R_\mathrm{S}$ and $R_\mathrm{out} = 3000\ R_\mathrm{S}$. The effective temperature profile for the AGN disk embedded with sBHs is shown in the top right panel of Figure~\ref{fig1}. For the sake of comparison, we also obtain the temperature profile for a pure SSD (the top left panel of Figure~\ref{fig1}) by eliminating $Q^{+}_{\mathrm{sBHs}}$ in Eq.~\ref{equ:Teff-sBHs} and fixing $f_{\mathrm{sBHs}}=0$. We compare the effective temperature profiles for both cases in the bottom panel of Figure~\ref{fig1}. In the presence of sBHs, the AGN disk temperature is significantly hotter than the SSD in the outer regions. Hence, we expect that, for a given wavelength, the disk size of an AGN disk with sBHs is larger than that of a pure SSD. 

Microlensing observations can constrain the AGN disk sizes. Microlensing observations essentially measure the half-light radius \citep{Mortonson2005}. The half-light radius can be estimated as follows. We consider perfect blackbody radiation at each radius and a face-on viewing angle. The local monochromatic luminosity can be expressed as 
\begin{equation}
    dL_\mathrm{\nu,sBHs} = \pi B_\mathrm{\nu}(T_\mathrm{eff,sBHs})Rd\phi dR \\,
\end{equation}
where $B_\mathrm{\nu}(T_\mathrm{eff,sBHs})=2h\nu^3/(c^2e^{h\nu/kT_\mathrm{eff,sBHs}}-c^2)$ is the Plank function, and $h$, $\nu$ and $k$ are the Planck constant, frequency and the Boltzmann constant, respectively. The half-light radius $R_\mathrm{half,sBHs}$ at wavelength $\lambda (=c/\nu)$ satisfies
\begin{equation} \label{equ:Rhalf-sBHs}
    \int_{R_\mathrm{in}}^{R_\mathrm{half,sBHs}}\int_{0}^{2\pi}dL_\mathrm{\nu,sBHs}
    =\frac{1}{2}\int_{R_\mathrm{in}}^{R_\mathrm{out}}\int_{0}^{2\pi}dL_\mathrm{\nu,sBHs}.
\end{equation}
The half-light radius of a pure SSD ($R_\mathrm{half,SSD}$) can be calculated following the same methodology. The spectral energy distribution (SED) is $L_\mathrm{\nu,sBHs}=\int_{R_\mathrm{in}}^{R_\mathrm{out}}\int_{0}^{2\pi}dL_\mathrm{\nu,sBHs}$. 

\section{Results} \label{sec:results}
We now calculate the SEDs (Section~\ref{subsec:SED}) and half-light radii (Section~\ref{subsec:rhalf}) of a pure SSD and an SSD with sBHs. Model parameters are set as follows. We consider three SMBH masses, $10^7$, $10^8$, or $10^9\ M_\mathrm{\odot}$. The mass of each sBH $M_\mathrm{sBH}$ is fixed at $50\ M_\mathrm{\odot}$, but our results remain unchanged when other $M_\mathrm{sBH}$ values are assumed. The dimensionless accretion rate $\dot{m}_{\mathrm{tot}}=\dot{M}_{\mathrm{tot}}/\dot{M}_\mathrm{\bullet,Edd}$ is 0.3 \citep{Kollmeier2006}, where $\dot{M}_\mathrm{\bullet,Edd}$ is the Eddington accretion rate of SMBH. The values of $f_\mathrm{sBHs}$ are taken as 0.06, 0.08, 0.10, and 0.12, respectively. Based on these parameters, the sBH number calculated from Eq.~\ref{equ:sBHs_number} and is about $10^3$--$10^5$, consistent with previous studies \citep[e.g.,][]{Artymowicz1993, Rozyczka1995, Gilbaum2022}. The ratio of the total mass of sBHs to the SMBH mass is $f_\mathrm{sBHs}\dot{m}_\mathrm{tot}$, which is significantly less than one. Gas in the AGN disk is still dominated by the SMBH gravity or the self-gravity. We calculate emission at wavelengths from $1000\ \textrm{\AA}$ to $8000\ \textrm{\AA}$. 

\subsection{Spectral energy distribution} \label{subsec:SED}
\begin{figure}
    \centering
    \includegraphics{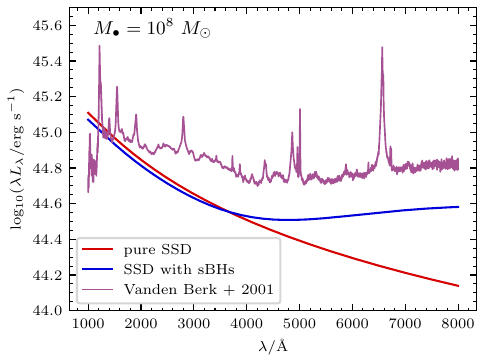}
    \includegraphics{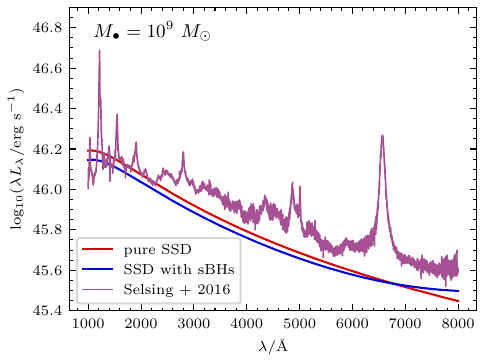}
    \caption{SEDs for the models and comparisons with composite quasar spectra. In each panel, the red curve is the SED for a pure SSD, the blue curve is the SED for an SSD embedded with a population of sBHs, and the purple curve is the composite quasar spectrum (normalized at $1450\ \textrm{\AA}$). The top panel is for $M_{\bullet}=10^8\ M_{\odot}$ with $f_{\mathrm{sBHs}}=0.1$ (for a pure SSD, $f_{\mathrm{sBHs}}\equiv 0$) and the composite quasar spectrum from \cite{Vanden_Berk2001}. When the wavelength is approximately larger than $4700\ \textrm{\AA}$, the SED of the SSD with sBHs is significantly larger than that of the pure SSD, and the spectral slope in this wavelength range is almost consistent with the composite spectrum. The bottom panel is for $M_{\bullet}=10^9\ M_{\odot}$ with $f_{\mathrm{sBHs}}=0.1$ and the composite quasar spectrum from \cite{Selsing2016}. In this case, the sBHs have weak effects on the SED.}
    \label{fig2}
\end{figure}

Previous studies have shown that star formation in the AGN accretion disk can revise the AGN SEDs \citep[e.g.,][]{Goodman2004, Thompson2005, Wang2023}, and the same is expected for accreting sBHs. The top panel of Figure~\ref{fig2} shows the SEDs of a pure SSD and an SSD embedded with sBHs for $M_{\bullet}=10^8\ M_{\odot}$ with $f_{\mathrm{sBHs}}=0.1$ (for a pure SSD, $f_{\mathrm{sBHs}}\equiv 0$). At wavelengths larger than $4700\ \textrm{\AA}$, the emission from the SSD embedded with sBHs is significantly larger than that from a pure SSD. This is because the main contribution of the sBHs is in the outer (long-wavelength emission) regions of the AGN accretion disk. At wavelengths shorter than $3600\ \textrm{\AA}$, the monochromatic luminosity from the SSD embedded with sBHs is slightly weaker than that from a pure SSD. This is simply because a fraction of gas is accreted to sBHs rather than the SMBH. The monochromatic luminosity from the accreting sBHs is equal to that from the AGN disk at a wavelength of $\sim 3600\ \textrm{\AA}$. For a more massive SMBH (e.g., $10^{9}\ M_{\odot}$; the bottom panel of Figure~\ref{fig2}), accreting sBHs have weak effects at these wavelengths. This is because that $T_\mathrm{eff,sBHs}\propto M_{\bullet}^{-1/4}\dot{m}_{\mathrm{tot}}^{1/4}$ according to Eqs. \ref{equ:qvis} and \ref{equ:Teff-sBHs}, i.e., $T_\mathrm{eff,sBHs}$ decreases with increasing $M_{\bullet}$ for fixing $\dot{m}_{\mathrm{tot}}$. As a result, the contribution due to accreting sBHs is prominent at wavelengths longer than $8000\ \textrm{\AA}$.

We compare the composite spectra from \cite{Vanden_Berk2001} and \cite{Selsing2016} with the SED of an SSD embedded with sBHs. We normalize the composite spectra to the SED of an SSD embedded with sBHs at $1450\ \textrm{\AA}$. The composite spectrum from \cite{Vanden_Berk2001} was compiled from a homogeneous sample of 2200 spectra with a median redshift of 1.25. Most spectra have $i$-band absolute magnitude fainter than $-24$ mag. Meanwhile, the composite spectrum of \cite{Selsing2016} was constructed from seven bright quasars with VLT/X-SHOOTER observations. Hence, we compare the composite spectrum of \cite{Vanden_Berk2001} with the model SED for $M_{\bullet}=10^8\ M_{\odot}$ with $f_{\mathrm{sBHs}}=0.1$ (the top panel of Figure~\ref{fig2}). In previous studies, the significant spectral slope change around $4700\ \textrm{\AA}$ is often attributed to the emission from hot dust \citep{Vanden_Berk2001}, host galaxy contamination \citep{Selsing2016}, or the diffuse emission from the broad-line region clouds \citep[e.g.,][]{Chelouche2019}. Interestingly, just like the composite spectrum of \cite{Vanden_Berk2001}, the SED of the SSD embedded with sBHs shows an evident shallow shape for wavelengths longer than $4700\ \textrm{\AA}$. The spectral slope of the SSD embedded with sBHs is almost identical to that of \cite{Vanden_Berk2001} at wavelengths larger than $4700\ \textrm{\AA}$. As mentioned above, this is because the main contribution of sBHs is in the outer regions of the AGN accretion disk. Our results suggest that this spectral slope variation may also be influenced by accreting sBHs in the AGN accretion disk \citep[see also][]{Sirko2003}. The composite spectrum of \cite{Selsing2016} compiled from bright AGNs is compared with the model SED of a $M_{\bullet}=10^{9}\ M_{\odot}$ with $f_{\mathrm{sBHs}}=0.1$ (the bottom panel of Figure~\ref{fig2}). In this case, accreting sBHs have weak effects on the model SED. Hence, the SED of an SSD embedded with sBHs does not show spectral shape change around $4700\ \textrm{\AA}$, just like the SED of a pure SSD and that of \cite{Selsing2016}. Note that our conclusions remain unchanged if we use other $f_{\mathrm{sBHs}}$. In summary, our model provides a new possible origin of the spectral shape variation around $4700\ \textrm{\AA}$ for less luminous AGNs. 

\subsection{Half-light radius} \label{subsec:rhalf}
\begin{figure}
    \centering
    \includegraphics{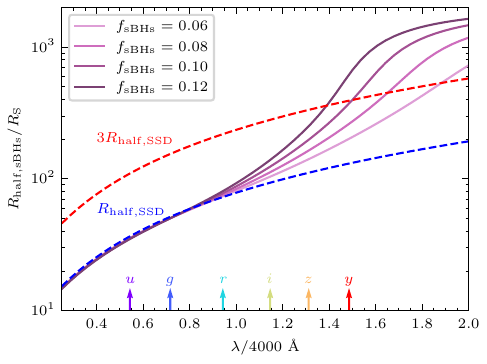}
    \caption{Half-light radii at different wavelengths for $M_{\bullet}=10^8\ M_{\odot}$. The x-axis is the rest-frame wavelength in units of $4000\ \textrm{\AA}$. The y-axis is the half-light radius for an SSD embedded with sBHs ($R_{\mathrm{half,sBHs}}$) and a pure SSD in units of $R_\mathrm{S}$. The purple curves represent the results with different $f_\mathrm{sBHs}$. The blue and red dashed curves are the half-light radius of a pure SSD ($R_{\mathrm{half,SSD}}$) and $3R_{\mathrm{half,SSD}}$, respectively. The colorful arrows indicate the rest-frame wavelengths of LSST bands at a redshift of $0.658$, i.e., the redshift of a low-redshift lensed quasar, RXS J113155.5-123155 \citep{Sluse2003}. The presence of accreting sBHs significantly alters the half-light radius-wavelength relation by increasing the half-light radius of the long-wavelength emission.}
    \label{fig3}
\end{figure}

Accreting sBHs modify the effective temperature distribution of the AGN accretion disk (Figure~\ref{fig1}), which can be distinguished by microlensing observations. The half-light radii at different wavelengths for an SSD embedded with sBHs can be calculated by Eq.~\ref{equ:Rhalf-sBHs}. Figure~\ref{fig3} shows the half-light radii at different wavelengths with $M_{\bullet}=10^8\ M_{\odot}$. For a given wavelength, the half-light radius increases as $f_\mathrm{sBHs}$ increases. The reason is that the larger $f_\mathrm{sBHs}$ is, the greater the contributions of the sBHs to the luminosity at large radii, resulting in a larger half-light radius. For the rest-frame wavelengths $\lesssim3600\ \textrm{\AA}$, the accreting sBHs have almost no influence on the half-light radius. This is because the main contribution of sBHs is at the outer regions of the AGN accretion disk, whose effective temperatures are too cold to produce short wavelength emission. If more sBHs are distributed in the short wavelengths radiation regions, the short wavelengths half-light radii will increase (see Section~\ref{subsec:4.1}). For rest-frame wavelengths larger than $5600\ \textrm{\AA}$, an SSD embedded with sBHs can yield half-light radii larger than three times that of a pure SSD, consistent with microlensing observations.

\begin{figure}
    \centering
    \includegraphics{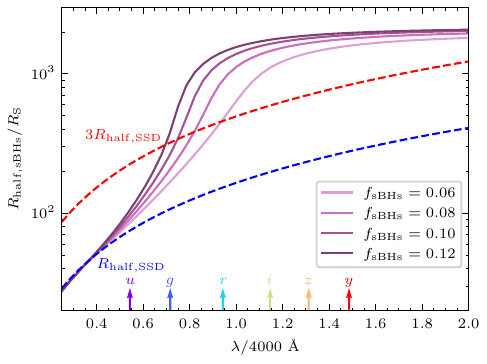}
    \caption{Same as Figure~\ref{fig3}, but for $M_{\bullet}=10^7\ M_{\odot}$. The wavelength ranges affected by accreting sBHs increase with decreasing SMBH mass.}
    \label{fig4}
\end{figure}

The range of wavelengths affected by accreting sBHs widens as the SMBH mass decreases (e.g., $10^{7}\ M_{\odot}$; Figure~\ref{fig4}). This is because the effective temperature of an SSD embedded with sBHs $T_\mathrm{eff,sBHs}\propto M_{\bullet}^{-1/4}\dot{m}_\mathrm{tot}^{1/4}(R/R_\mathrm{S})^{-3/4}(1+\epsilon)^{1/4}$ according to Eqs. \ref{equ:qvis} and \ref{equ:Teff-sBHs}. For the same $R/R_\mathrm{S}$ and $\dot{m}_\mathrm{tot}$, $T_\mathrm{eff,sBHs}$ increases with decreasing $M_{\bullet}$. As a result, the contribution of accreting sBHs in outer regions can lead to short wavelength emission increases with a decrease in the SMBH mass. With our hypothetical sBH distribution, the accretion disk embedded with sBHs of an SMBH with $M_{\bullet}=10^7\ M_{\odot}$ has half-light radii three times or more of a pure SSD for wavelengths longer than $\sim 3000\ \textrm{\AA}$. 

The relationship between the half-light radius and wavelength is significantly altered by accreting sBHs in the AGN accretion disk. The Legacy Survey of Space and Time \citep[LSST;][]{Ivezic2019} and the Wide Field Survey Telescope \citep[WFST;][]{Wang2023-WFST} will perform high resolution multi-band surveys in the south and north sky, respectively. They can measure the half-light radii of some lensed quasars in multiple bands, which can verify our model calculations and probe the distributions of accreting sBHs in AGN accretion disks.

\section{Discussions} \label{sec:discussion}
\subsection{Stellar black hole distributions} \label{subsec:4.1}
\begin{figure*}
    \centering
    \includegraphics{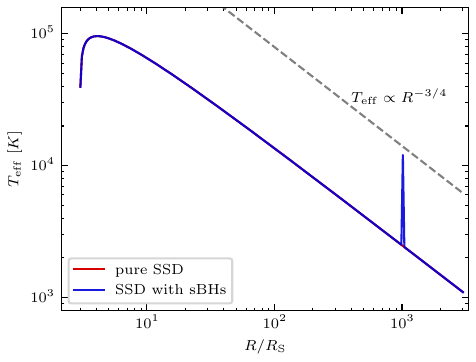}
    \includegraphics{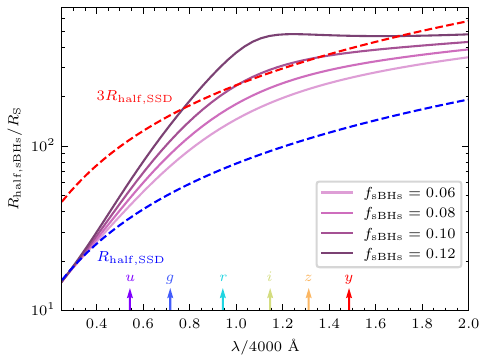}
    \caption{The left panel is same as the bottom panel of Figure \ref{fig1} and the right panel is same as Figure~\ref{fig3}, but with all sBHs distributed at $R_\mathrm{sBHs}=1000\ R_\mathrm{S}$. For this sBHs distribution, the effects of sBHs are on the same radius, and only the effective temperature at $1000\ R_\mathrm{S}$ is changed. This distribution of sBHs allows the AGN accretion disk to have large half-light radii even at short wavelengths (i.e., $2000\ \textrm{\AA}$).}
    \label{fig5} 
\end{figure*}

Accreting sBHs have significant effects on the SEDs and half-light radii of long-wavelength optical emission. Although we fix $R_\mathrm{out}$ at $3000\ R_\mathrm{S}$, it is possible that some sBHs lacate at $> 3000\ R_\mathrm{S}$ and can also have a significant impact on the SEDs and half-light radii at infrared bands. Meanwhile, similar to planets in stellar disks, sBHs can interact with gas in the AGN accretion disk and migrate inward or outward. Detailed calculations suggest that the disk can form several migration traps at which migrating sBHs are trapped and accumulated \citep[e.g.,][]{Bellovary2016, Grishin2023}. \cite{Grishin2023} note that the migration traps are approximately at $10^{3}$--$10^{5}$ gravitational radii. Thus, we consider a case where all sBHs are trapped at the same radius of an AGN accretion disk. Figure~\ref{fig5} shows the results for all sBHs distributed at a radius of $1000\ R_\mathrm{S}$ for an AGN disk with $M_{\mathrm{BH}}=10^{8}\ M_{\odot}$. As expected, sBHs now only affect the effective temperature at $1000\ R_\mathrm{S}$. The corresponding half-light radii are larger than a pure SSD even on wavelengths as short as $2000\ \textrm{\AA}$. This is because when all sBHs are at $1000\ R_\mathrm{S}$, they are closer to the radiation region of the short wavelengths than the sBH distribution of Eq.~\ref{equ:NR}. As a result, these sBHs have a larger impact on the half-light radii of short wavelength emission than the sBH distribution of Eq.~\ref{equ:NR}.

Real sBH distributions, which are likely to be much more complex than we have explored, can have various impacts on the half-light radii and the effects can be understood as follows. Consider a general sBH distribution of $N(R)\propto (R/R_{\mathrm{S}})^{\beta}$ with $\beta=2$ is identical to the sBH distribution of Eq.~\ref{equ:NR}. The ratio of the sBH heating rate to the local viscous heating rate (Eq.~\ref{equ:ratio}) $\epsilon \propto (R/R_{\mathrm{S}})^{\beta+1}$. According to Eqs. \ref{equ:qvis} and \ref{equ:Teff-sBHs}, the corresponding effective temperature $T_\mathrm{eff,sBHs}\propto M_{\bullet}^{-1/4}\dot{m}_\mathrm{tot}^{1/4}(R/R_\mathrm{S})^{-3/4}(1+\epsilon)^{1/4}$. As $\beta$ decreases (i.e., more sBHs are distributed in the inner regions), there are two consequences. First, the critical radius ($R_{\mathrm{c}}$) at which $\epsilon = 1$ moves inward. The additional heating due to sBHs only plays an important role in the disk regions outside $R_{\mathrm{c}}$. As $R_{\mathrm{c}}$ moves inward, the shorter-wavelength ($\sim \lambda_{\mathrm{c}}\equiv hc/(kT_{\mathrm{eff}}(R_{\mathrm{c}}))$) emission regions are also affected by sBHs. As a result, the half-light radii at short wavelengths ($\simeq \lambda_{\mathrm{c}}$) can increase. Second, the absolute value of $\epsilon$ is smaller in outer regions . As a result, the half-light radii at long wavelengths ($\gg \lambda_{\mathrm{c}}$) decrease. Hence, the tendency of the half-light radius with wavelength is a prospective method for inferring the sBH distribution. Inferring the sBH distribution requires long rest-frame wavelength half-light radius measurements. RXS J113155.4-123155 is one of the nearest gravitationally lensed quasars found to date, with a redshift of 0.658 \citep{Sluse2003} and $M_{\bullet}=10^{7.78}\ M_{\odot}$ \citep{Peng2006}. The arrows in Figure~\ref {fig3}, \ref{fig4}, and \ref{fig5} represent the rest-frame wavelengths of this source probed by LSST filters. The LSST is able to obtain the half-light radii in the $rizy$ bands that are significantly influenced by sBHs. 

Collisions and mergers of sBHs in an AGN accretion disk are a possible source of gravitational waves. For an AGN with $M_{\bullet}=10^{8}\ M_{\odot}$ and $f_\mathrm{sBHs}=0.1$, the maximum number of sBHs in the zones at $3000\ R_\mathrm{S}$ is 24 for the sBH distribution of Eq.~\ref{equ:NR}. \cite{Whitehead2023} show that two sBHs will form a binary when their radial distance is $[1.85,\ 2.4]r_\mathrm{H}$, where $r_\mathrm{H}$ is the Hill radius which is comparable to the zone size we considered in Section \ref{sec:Model}. The results of \cite{Whitehead2023} are obtained for an AGN accretion disk with $M_{\bullet}=4\times10^{6}\ M_{\odot}$. The merger may become more difficult for a more massive SMBH accretion disk as the AGN disk gas density is lower. Nevertheless, we speculate that sBHs in each zone of the accretion disk can merge to form a massive sBH. We stress that our results for the SEDs and half-light radii are independent of the masses of sBHs.

\subsection{AGN accretion physics}\label{subsec:4.2}
The quasar accretion disk sizes measured by microlensing are 2--4 times larger than the SSD model predictions \citep[e.g.,][]{Morgan2010}. The accretion disk sizes predicted by the SSD model depend on the assumptions of blackbody radiation and $T_\mathrm{eff}\propto R^{-3/4}$. Therefore, one way to solve this problem is to flatten the disk temperature profile. Several models have been proposed to explain the differences between microlensing observations and the SSD model, such as the inhomogeneous disk \citep{Dexter2011}, non-blackbody disks \citep{Hall2018}, and the windy disk models \citep{Li2019, Sun2019-wind}. 

Here, we propose a new possibility. Accreting sBHs can flatten the effective temperature distribution of the accretion disk because the effective temperature in the outer regions of the accretion disk increases (Figure~\ref{fig1}). It should be noted that sBHs mainly affect the half-light radii of relatively long wavelengths, and the affected wavelength ranges depend upon the sBH distribution. Alternative mechanisms should play dominant roles for very short wavelengths (e.g., $1500\ \textrm{\AA}$). The inhomogeneous disk model proposed by \cite{Dexter2011} states that the local temperature of the accretion disk is perturbed due to MHD turbulence and returns to a mean value within a characteristic time. The temperature variations in the inner regions of the accretion disk should be more significant than the outer regions because the characteristic time is expected to be positively correlated with radius \citep[see, e.g.,][]{Sun2020-CHAR, Zhou2024}. Thus, it is natural to expect that the half-light radii at short and long wavelengths are affected by the disk temperature fluctuations and accreting sBHs, respectively.

Embedded sBHs have significant effects on SED for SMBHs with $M_{\bullet}\leq 10^{8}\ M_{\odot}$ (Figure~\ref{fig2}). Hence, they can affect the bolometric correction of the $5100\ \textrm{\AA}$ luminosity, making the bolometric correction mass-dependent. 

The additional heating due to sBHs also reduces the viscous timescale. In the outer regions, the viscous timescale of the AGN SSD model can be even much longer than the AGN lifetime. \cite{Gilbaum2022} propose a possible solution, i.e., the additional heating due to sBHs increases the gas temperature and significantly reduces the viscous timescale. 

If sBHs are distributed in the optically thin regions \citep[$\sim 10^4\ R_\mathrm{S}$; e.g.,][]{Thompson2005}, the X-ray photons produced by sBHs can escape instead of being reprocessed by the accretion-disk gas and contribute to the observed AGN X-ray emission. We calculate the soft X-ray luminosity at $2\ \mathrm{keV}$, $L_\mathrm{2keV}$, for an SMBH with a mass of $10^8\ M_{\odot}$ and $\dot{m}$ of $0.3$, and an sBH with a mass of $50\ M_{\odot}$, accreting in the Eddington limit. For the SMBH accretion, we obtain the monochromatic luminosity at $2500\ \textrm{\AA}$, $L_\mathrm{2500\textrm{\AA}}$, using the SSD model. With the $L_\mathrm{2keV}$--$L_\mathrm{2500\textrm{\AA}}$ relation \citep[Figure 3 in][]{Lusso2016}, we derive $L_\mathrm{2keV}$ for the SMBH accretion as $4.9\times 10^{43}\ \mathrm{erg\ s^{-1}}$. For an sBH accretes at the Eddington limit, the major soft X-ray emission is considered as thermal radiation from the sBH accretion disk \citep[e.g.,][]{Done_2007}. Thus, $L_\mathrm{2keV}$ for an sBH can also be estimated using the SSD model, which is $1.6\times 10^{39}\ \mathrm{erg\ s^{-1}}$. If there are abundant sBHs in the optically thin regions, these sBHs may contribute considerably to the observed AGN soft X-ray emission.

\subsection{Implications for strong lensing time-delay cosmography}
The time delays between different images of strong gravitational lensing are used to constrain the cosmological model. Still, the accuracy of the time delay measurement is limited due to microlensing \citep[e.g.,][]{Bonvin2017, Tie2018}. Microlensing has different magnifications for the different disk regions, and the observations are a superposition of the different regions. This causes fluctuations in the time delays between different images of strong gravitational lensed quasars. \cite{Tie2018} conservatively estimate that microlensing will affect strong lensing time delays at $\sim$ days based on the SSD. If there are a large number of sBHs in the outer region of the AGN accretion disk, the size can be up to five times or more than SSD for long wavelengths (see Figures~\ref{fig3} and \ref{fig4}). The effect of microlensing on strong lensing time delays will be significantly increased.

\section{Summary} \label{sec:Summary}
We have explored the SED and half-light radius of an AGN accretion disk embedded with accreting sBHs. We have shown that a population of accreting sBHs residing in the outer regions of an AGN accretion disk can produce observable features in SEDs and microlensing observations. The main conclusions can be summarized as follows.
\begin{enumerate}
\item Embedded sBHs can cause the effective temperature of the outer regions to be significantly higher than that of the pure SSD. (see Section~\ref{sec:Model}; Figure~\ref{fig1}).
\item Compared to a pure SSD, an SSD embedded with sBHs produces a redder SED, which may contribute significantly to the spectral slope change around $5000\ \textrm{\AA}$ for AGNs with low ($\lesssim 10^8\ M_{\odot}$) SMBH masses (see Section~\ref{subsec:SED}, Figure~\ref{fig2}).
\item The dependence of the half-light radius with wavelength for an SSD embedded with sBHs significantly differs from that of a pure SSD. With suitable sBH distributions, the model half-light radius is consistent with microlensing observations (see Sections~\ref{subsec:rhalf} and \ref{subsec:4.1}; Figures~\ref{fig3}, \ref{fig4} and \ref{fig5}).
\item The dependence of the half-light radius with wavelength can probe the sBH distribution in the AGN accretion disk (see Sections~\ref{subsec:rhalf} and \ref{subsec:4.1}; Figures~\ref{fig3} and \ref{fig5}).
\end{enumerate}

As mentioned in Section~\ref{subsec:4.2}, the primary influence of sBHs is at the outer regions of the AGN accretion disk; the inner regions may be more dominated by other mechanisms, such as the inhomogeneous disk proposed by \cite{Dexter2011}. Therefore, we will consider the accretion disk size in conjunction with sBHs and other models in future works.

\begin{acknowledgments}
We thank the referee for his/her constructive comments that improve the manuscript. We acknowledge support from the National Key R\&D Program of China (No.~2023YFA1607900, No.~2023YFA1607903, No.~2023YFA1608100). S.Y.Z. and M.Y.S. acknowledge support from the National Natural Science Foundation of China (NSFC-12322303), and the Natural Science Foundation of Fujian Province of China (No.~2022J06002). J.M.W. acknowledges support from the National Natural Science Foundation of China (NSFC-11991050, NSFC-12333003). J.X.W. acknowledges support from the National Natural Science Foundation of China (NSFC-12033006, NSFC-12192221). Y.Q.X. acknowledges support from the National Natural Science Foundation of China (NSFC-12025303, NSFC-12393814) and the Strategic Priority Research Program of the Chinese Academy of Sciences (grant NO.~XDB0550300). 
\end{acknowledgments}

\vspace{5mm}

\software{Matplotlib \citep{matplotlib}, Numpy \citep{2020NumPy-Array}, Scipy \citep{2020SciPy-NMeth}}

\bibliography{ref.bib}{}
\bibliographystyle{aasjournal}

\end{document}